\pdfoutput=1

\documentclass[11pt]{article}
\usepackage{listings}
\usepackage{amsmath}
\usepackage{ACL2023}
\usepackage{graphicx}
\usepackage{booktabs}
\usepackage{subcaption}
\usepackage{amsmath}
\usepackage{algorithm}
\usepackage{algpseudocode}
\usepackage{latexsym}
\usepackage{makecell}
\usepackage{textcomp}
\usepackage{tabularx} 
\usepackage{times}
\usepackage{latexsym}
\usepackage{multirow}
\usepackage{amssymb}
\usepackage{amsmath}

\usepackage[T1]{fontenc}

\usepackage[utf8]{inputenc}
\usepackage{float}

\usepackage{microtype}

\usepackage{inconsolata}

\title{SpeechT-RAG: Reliable Depression Detection in LLMs with Retrieval-Augmented Generation Using Speech Timing Information}

\author{Xiangyu Zhang$^{1}$, \textbf{Hexin Liu}$^{2}$\thanks{\ \ Corresponding author.}, \textbf{Qiquan Zhang}$^{1}$, \textbf{Beena Ahmed}$^{1}$, \textbf{Julien Epps}$^{1}$\\
The University of New South Wales$^1$,
Nanyang Technological University$^2$}
\begin{document}
\maketitle

\begin{abstract}
Large Language Models (LLMs) have been increasingly adopted for health-related tasks, yet their performance in depression detection remains limited when relying solely on text input. While Retrieval-Augmented Generation (RAG) typically enhances LLM capabilities, our experiments indicate that traditional text-based RAG systems struggle to significantly improve depression detection accuracy. This challenge stems partly from the rich depression-relevant information encoded in acoustic speech patterns — information that current text-only approaches fail to capture effectively. To address this limitation, we conduct a systematic analysis of temporal speech patterns, comparing healthy individuals with those experiencing depression. Based on our findings, we introduce Speech Timing-based Retrieval-Augmented Generation, SpeechT-RAG, a novel system that leverages speech timing features for both accurate depression detection and reliable confidence estimation. This integrated approach not only outperforms traditional text-based RAG systems in detection accuracy but also enhances uncertainty quantification through a confidence scoring mechanism that naturally extends from the same temporal features. Our unified framework achieves comparable results to fine-tuned LLMs without additional training while simultaneously addressing the fundamental requirements for both accuracy and trustworthiness in mental health assessment.
\end{abstract}

\section{Introduction}
\begin{figure}[t]
    \centering
    \includegraphics[width=0.5\textwidth]{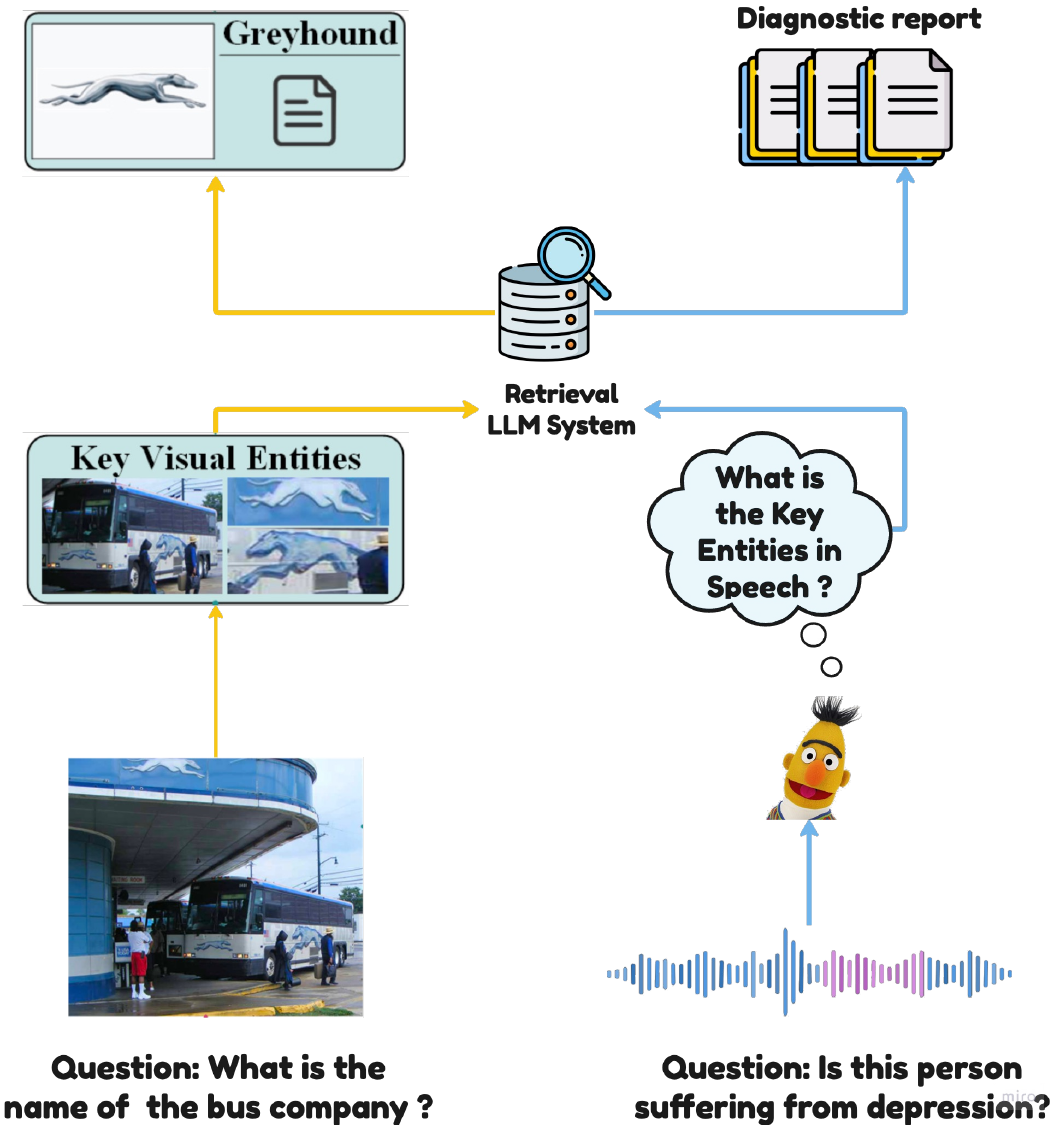}
    \caption{Motivation for Speech Timing-based RAG. While visual RAG systems can leverage distinct entities as retrieval keys (left), identifying analogous "key entities" in speech for depression detection is challenging.}
    \vspace{-16pt}
    \label{fig:motivation}
\end{figure}

\begin{figure*}[t]
    \centering
    \includegraphics[width=1\textwidth]{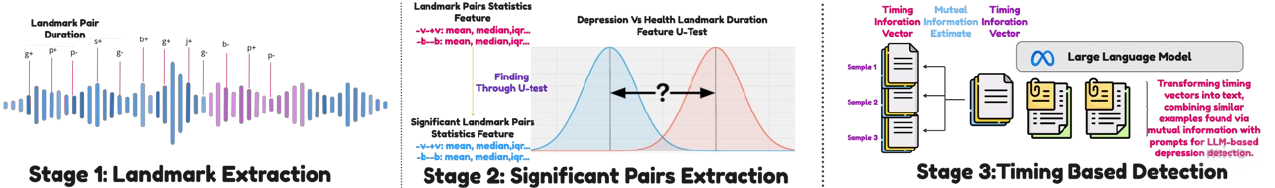}
    \caption{Overview of method: (1) extract acoustic landmarks and encode temporal information using durations between consecutive bigram pairs; (2) identify statistically significant landmark pairs that differentiate depression from health based on duration features; and (3) utilize timing keys and text-based representations for RAG.}
    \vspace{-12pt}
    \label{fig:overview}
\end{figure*}
Large Language Models (LLMs)\cite{brown2020language,achiam2023gpt} have been extensively utilized in health-related applications, achieving notable success in tasks such as medical evidence summarization\cite{tang2023evaluating}, supporting decision-making in general surgery~\cite{oh2023chatgpt}, and assisting in gastroenterological diagnoses~\cite{lahat2023evaluating}. These models have demonstrated exceptional capabilities in transforming traditional healthcare workflows by efficiently processing complex medical data and enhancing decision-making processes. However, in specialized domains like depression detection, LLMs based solely on text have shown limited effectiveness~\cite{ohse2024zero,zhang2024llms,wu2023holistic}, often requiring extensive task-specific fine-tuning and large annotated datasets to achieve satisfactory results. This reliance on significant training resources makes their application to depression detection both time-consuming and costly.

When the performance of LLMs is restricted, Retrieval-Augmented Generation (RAG) is often employed as an alternative. By allowing LLMs to retrieve task-specific information from external sources during inference, RAG has shown promise in mitigating the need for extensive fine-tuning across many tasks~\cite{guu2020retrieval,lewis2020retrieval}. However, in the context of depression detection, the limited performance of RAG highlights the inherent constraints of text-based approaches shown in our experiment. These limitations stem from text's inability to capture critical non-verbal cues, such as tone, prosody, and timing, which play a significant role in identifying depression~\cite{williamson2013vocal,quatieri2012vocal}. Therefore, it is crucial to address these gaps by seeking additional sources of information that can complement text-based methods and overcome their inherent limitations.

Numerous studies have demonstrated that acoustic features in speech, particularly temporal and prosodic patterns, contain rich information highly relevant to depression detection, and many existing works have successfully leveraged these acoustic characteristics as the primary modality for this task~\cite{huang2019investigation,zhao2020hierarchical}. However, directly integrating acoustic features into text-based LLMs has often been detrimental to their performance~\cite{zhang2024multimodal,zhang2024llms}, making multimodal RAG a promising alternative. In computer vision, RAG systems frequently utilize salient visual elements or regions of interest (such as objects or scene segments) as input to retrieve relevant information~\cite{jian2024large,xia2024rule}, but an analogous technique for identifying and utilizing key acoustic segments is lacking in the speech domain.~Furthermore, speech-enabled LLMs typically rely on encoders capable of processing only up to 30 seconds of audio~\cite{efficient_lid, chu2023qwen,radford2023robust, liu2024aligning}, which limits their applicability for tasks requiring long-term context~\cite{zhang2024mamba,chen2025selective,zhang2024speaking}. This constraint poses a significant challenge for depression detection, as the task often necessitates analyzing extended acoustic patterns across entire conversations~\cite{zhang2024llms,sun2022tensorformer}.~Given these limitations, directly using speech LLMs is impractical for this domain, prompting the need to explore alternative approaches that leverage the advantages of text-based LLMs while effectively incorporating acoustic information from speech.

Building on the aforementioned limitations, a key research question emerges: 

\textit{How can acoustic temporal patterns from speech be integrated into text-based LLMs for depression detection without training the LLMs?}

To address this question, it is essential to develop a speech-based RAG framework. The central challenge lies in identifying the "region of interest" within acoustic patterns that can act as the optimal key for the RAG process. This region must encapsulate the distinguishing temporal and prosodic features that separate individuals with depression from healthy individuals, enabling the retrieval and generation components to better leverage acoustic information and achieve robust, scalable performance in depression detection. Prior research has established that acoustic landmarks provide critical features for depression detection systems, with particular emphasis on the discriminative power of their temporal patterns in differentiating between individuals with and without depression.~\cite{zhang2024llms,huang2019investigation,huang2019natural,huang2019speech}. Building on this insight, we designed the Speech Timing-based Retrieval-Augmented Generation framework, SpeechT-RAG, which leverages acoustic temporal patterns from landmarks to integrate speech information into text-based LLMs for depression detection without requiring additional training.

In developing our framework, we observed that these acoustic temporal patterns not only serve as effective features for depression detection but also naturally encode prediction reliability. While traditional methods for generating confidence scores often face challenges with LLMs~\cite{yona2024can,wu2024confidence}, the temporal patterns we identify through acoustic landmarks provide an interpretable basis for assessing prediction reliability. This insight enables us to develop an integrated approach where temporal information drives both detection and uncertainty quantification, enhancing system reliability in contexts where trustworthy decision-making is essential.

\section{Preliminary}
\begin{table}[t]
\centering
\caption{Descriptions of the seven acoustic landmarks used in this study~\cite{liu1996landmark,zhang2024llms}.}
\begin{tabular}{l|p{0.65\linewidth}}
\hline
\hline
\textbf{Landmark} & \textbf{Description} \\
\hline
\hline
\textbf{g} & Onset (+) or offset (–) of vocal fold vibrations. \\ 
\textbf{b} & Onset (+) or offset (–) of turbulent noise during obstruent regions. \\ 
\textbf{s} & Onset (+) or offset (–) of nasal releases or closures. \\ 
\textbf{v} & Onset (+) or offset (–) of voiced frication. \\ 
\textbf{p} & Onset (+) or offset (–) of periodicity in voiced patterns. \\ 
\textbf{f} & Onset (+) or offset (–) of frication in unvoiced patterns. \\ 
\textbf{j} &  Abrupt upward jump (+) or abrupt downward jump (–) in F0 \\ 
\hline 
\hline
\end{tabular}
\label{tab:acoustic_landmarks}
\end{table}
\subsection{Acoustic Landmark}
The concept of acoustic landmarks originates from studies on distinctive features~\cite{garvin1953preliminaries,zhang2024auto}, emphasizing their role in phonetic contrasts and speech comprehension. Researchers have proposed that listeners rely on acoustic landmarks to extract essential cues for interpreting distinctive features, underscoring their significance in auditory processing~\cite{liu1996landmark}. Over the years, acoustic landmarks have been extensively explored in various domains, including speech recognition~\cite{liu1996landmark,he2019ctc}, and more recently, in the field of depression detection~\cite{huang2018depression,huang2019investigation,zhang2024llms,zhang2025pre}, where they have shown potential for identifying speech patterns indicative of mental health conditions. Recent studies further suggest that the temporal information embedded within acoustic landmarks may hold particular value for applications in the health domain~\cite{ishikawa2017toward,huang2019investigation}.

\subsection{Computer Vision Retrieval-Augmented Generation}

As illustrated in Figure~\ref{fig:motivation}, a common approach in computer vision for RAG involves extracting key entities or regions of interest from images~\cite{jian2024large,xia2024rule}, such as objects or salient areas, to serve as keys for retrieval and inputs for generation. This process effectively identifies distinctive features between images, enabling models to focus on task-relevant variations and improve interpretability. In contrast, speech lacks well-defined structures that can serve as analogous keys for RAG. Unlike visual data, where spatial features are clear, existing speech studies primarily rely on frame-level features or aggregated embeddings~\cite{wu2023self,wu2024confidence}, which often overlook nuanced temporal dynamics crucial for tasks like depression detection~\cite{dineley2024variability}. This gap highlights the need for innovative methods to define and extract meaningful keys from speech that align with the RAG paradigm.

\section{SpeechT-RAG Framework}
Our methodology involves three steps: First, we extract acoustic landmarks and construct sequential bigram pairs, embedding temporal information through their durations. Second, we identify statistically significant landmark pairs that exhibit notable differences between the two groups based on their duration statistics. Finally, we leverage these timing keys and text-based representations of temporal information to perform RAG with LLMs.

\subsection{Landmark Extraction and Consecutive Bigram Construction}
Previous studies have shown that temporal information is crucial for effective depression detection, as timing and rhythm often exhibit distinct patterns in affected individuals~\cite{huang2019investigation,huang2019natural,dineley2024variability}. To capture this essential temporal information, we leverage acoustic landmarks, which serve as indicators of key acoustic changes in speech, such as transitions between voiced and unvoiced regions, the onset of bursts, or sustained energy patterns in syllabic regions~\cite{liu1996landmark,boyce2012speechmark,zhang2024auto}. By utilizing the durations between consecutive acoustic landmarks, we extract the intrinsic temporal features embedded within speech signals, enabling a more precise analysis of timing patterns critical for identifying depression-specific characteristics.

Figure~\ref{fig:overview} stage one illustrates the process of extracting acoustic landmarks from speech, with a primary focus on the temporal intervals—\textit{durations}—between consecutive landmarks as the core feature of interest. Table~\ref{tab:acoustic_landmarks} lists the specific landmarks employed in this study, including \textbf{glottal (g)}, indicating vocal fold vibrations; \textbf{burst (b)}, representing plosive events; \textbf{syllabic (s)}, capturing vowels and sustained sonorants; \textbf{frication (f)} and \textbf{voiced frication (v)}, identifying unvoiced and voiced fricatives; and \textbf{periodicity (p)}, denoting recurring voiced patterns. \textbf{Jump (j)}, denotes abrupt change in F0. These landmarks collectively provide a comprehensive framework for analyzing the temporal dynamics of speech.

To encode the temporal relationships, we construct \textit{landmark bigrams} by pairing consecutive landmarks within each utterance~\cite{huang2019investigation,huang2019natural}. For two sequential landmarks \( l_i \) and \( l_{i+1} \), the bigram is defined as \( b_{i \rightarrow i+1} = (l_i, l_{i+1}) \), and the corresponding duration is computed as:
\begin{equation}
d_{i \rightarrow i+1} = t(l_{i+1}) - t(l_i),
\end{equation}
where \( t(l_i) \) represents the timestamp of landmark \( l_i \). This duration captures the temporal interval between two adjacent distinctive events, offering insights into speech timing. Unlike frame-based representations, this approach emphasizes the natural sequence and timing of events, characterizing the rhythmic properties of acoustic speech.

To analyze these durations, we calculated an enhanced set of statistical features for each landmark bigram. While previous work~\cite{huang2019investigation} utilized basic statistics like mean and standard deviation, we introduce the interquartile range (IQR) to better capture the robustness of temporal variations, as IQR is less sensitive to outliers that commonly occur in natural speech patterns. Our statistical feature set includes the \textit{mean}, \textit{median}, \textit{standard deviation}, \textit{interquartile range (IQR)}, \textit{minimum}, and \textit{maximum} values:
\begin{equation}
\text{Statistical} = \{\mu, \text{med}, \sigma, \text{IQR}, \text{min}, \text{max}\}.
\end{equation}
These features provide a detailed characterization of the temporal patterns in speech, preparing for distinguishing between healthy individuals and those with depression.

\subsection{Identifying Distinctive Landmark Pairs}
\begin{table}[t]
\centering
\setlength{\tabcolsep}{4pt} 
\renewcommand{\arraystretch}{1.2} 
\caption{Top 5 landmark pairs with the lowest \textit{p}-values}
\label{tab:top_landmark_pairs}
\scriptsize
\begin{tabular}{@{}lcccc@{}}
\hline
\hline
\textbf{Landmark Pair} & \textbf{\( U \)-Statistic} & \textbf{\( p \)-Value} & \textbf{Health \(\mu\)} & \textbf{Depression \(\mu\)} \\
\hline
\( +s--v \) & 23042.5 & 0.00078 & 0.0795 & 0.0991 \\
\( +j--v \) & 33793.5 & 0.00112 & 0.0309 & 0.0197 \\
\( -v-+j \) & 43035.5 & 0.00193 & 0.0725 & 0.0548 \\
\( -v--j \) & 30640.5 & 0.01029 & 0.0523 & 0.0670 \\
\( +g--v \) & 32691.0 & 0.01097 & 0.0100 & 0.0088 \\
\toprule
\end{tabular}
\end{table}

To identify the landmark pairs that exhibit statistically significant differences between healthy individuals and those with depression, We utilized the Mann-Whitney U test~\cite{mcknight2010mann} to systematically identify landmark bigram pairs whose duration distributions differ significantly between the non-depressed and depression groups. For a given bigram pair \( b_{i \rightarrow i+1} = (l_i, l_{i+1}) \), its aggregated duration set \( D_{b_{i \rightarrow i+1}} \) consists of durations calculated across all samples. Specifically, we computed the set of statistical features \( \{ \text{mean}, \text{variance}, \text{interquartile range}, \dots \} \) for \( D_{b_{i \rightarrow i+1}} \). The Mann-Whitney U test was then applied to compare the feature distributions between the health and depression groups for each bigram pair:
\begin{equation}
U, p = \text{MannWhitneyU}(D^\text{health}_{b_{i \rightarrow i+1}}, D^\text{depression}_{b_{i \rightarrow i+1}})
\end{equation}
where \( p \)-values below 0.05 indicate significant differences in the temporal patterns of the bigram pair between the two groups.

Table~\ref{tab:top_landmark_pairs} highlights the top five landmark pairs with the lowest \( p \)-values, indicating significant differences in their durations between the health and depression groups. These pairs, such as \( +s--v \) and \( +j--v \), demonstrate how temporal dynamics captured by landmark bigrams can effectively distinguish between the two groups. For subsequent tasks, we focus exclusively on those landmark bigrams with \( p \)-values less than 0.05, utilizing their statistical features to represent depression-specific characteristics.

\subsection{SpeechT-RAG Framework Implementation}
To effectively integrate speech timing information into the LLM for depression detection, we leverage a Speech Timing-Based Retrieval-Augmented Generation (RAG) framework as shown in Figure~\ref{fig:overview} stage 3. This framework selects representative examples from the training set based on mutual information (MI) scores and incorporates their temporal information into the text-based LLM.

MI measures the dependency between two random variables \(X\) and \(T\). The MI for the random variable \(X \in \mathbb{R}^{L \times D}\) and random variable \(T \in \mathbb{R}^{L \times D}\) is expressed as:

\small
\begin{equation}
\label{MI-Equation}
\begin{aligned}
I_i(X;T_i) &= H(X) - H(X \mid T_i) \\
       &= D_{KL}\left(P(X,T_i) \parallel P(X) \otimes P(T_i)\right)
\end{aligned}
\end{equation}
\normalsize
where \(H(X)\) is the entropy of \(X\) and \(H(X \mid T)\) is the conditional entropy of \(X\) given \(T\),~$D_{KL}$ denotes KL-divergence.~Since directly estimating \(I(X; T)\) is computationally intractable, MINE~\cite{belghazi2018mutual} approximates MI using a deep neural network, an approach that has found wide application across various domains~\cite{ravanelli2020multi,zhang2024rethinking}:

\small
\begin{equation}
I_\Theta(X; T) = \mathbb{E}_{P(X,T)}[\psi_\theta] - \log(\mathbb{E}_{P(X)P(T)}[e^{\psi_\theta}])
\end{equation}
\normalsize

where \(\psi_\theta\) is a statistics network parameterized by \(\theta\). Gradients of \(\psi_\theta\) are computed by random batch sampling, ensuring efficient estimation of \(I(X; T)\).

For each training sample \(\mathbf{x}_{\text{train}}\), we compute its MI with a test sample \(\mathbf{x}_{\text{test}}\), resulting in a set of MI scores. To ensure robustness, we calculate the average MI across test samples as:
\begin{equation}
\bar{I}(\mathbf{x}_{\text{train}}) = \frac{1}{M} \sum_{j=1}^{M} I(\mathbf{x}_{\text{test}}^j, \mathbf{x}_{\text{train}})
\end{equation}
where \(M\) is the number of test samples. Based on these MI scores, the top \(n\) training examples most similar to the test sample are selected, including \(n\) examples each from the health and depression classes:
\begin{equation}
\mathcal{R} = \mathcal{H} \cup \mathcal{D}, \quad |\mathcal{H}| = |\mathcal{D}| = n,
\end{equation}
where \(\mathcal{H}\) and \(\mathcal{D}\) denote the selected health and depression examples, respectively.

To incorporate timing information into the LLM, the temporal features (e.g., mean, variance) of landmark bigram durations are converted into a structured text format. Each timing sequence is formatted as:
\begin{equation}
\mathbf{t}_{\text{sample}} = \text{Format}(\mathbf{d}_{\text{bigram}})
\end{equation}
where \(\mathbf{d}_{\text{bigram}}\) represents the statistical features of bigram durations. For example:
\begin{equation}
\mathbf{t}_{\text{sample}}: \, \text{+s--v (mean: 0.08, var: 0.01)},
\end{equation}
Here, \textit{+s--v} represents a transition from a syllabic onset to a voiced frication.

The formatted representations of the retrieved examples are concatenated with the timing information of the test sample to construct a prompt:
\begin{equation}
\mathbf{P} = \mathbf{t}_{\text{example}_1} + \mathbf{t}_{\text{example}_2} + \dots + \mathbf{t}_{\text{test}}
\end{equation}
where \(+\) denotes concatenation. The prompt, along with an instruction, is fed into the LLM for classification. The LLM generates predictions for the test sample as either \textit{Health} or \textit{Depression}. 

\section{Confidence Score Estimation}
Confidence estimation is a critical component of health-related AI systems, offering a measure of reliability for predictions, which is essential for clinical decision-making~\cite{edin-etal-2024-unsupervised,kang-etal-2024-cure}. While traditional machine learning systems frequently employ confidence scoring techniques~\cite{wu2024confidence}, their application to LLMs remains challenging~\cite{yona-etal-2024-large,chaudhry2024finetuning}. To address this, we propose a framework based on Gaussian Process Classifiers (GPCs) that leverages speech timing information, specifically the temporal features of landmark bigrams with \( p \)-values below 0.05, to predict confidence scores.

\paragraph{Gaussian Process Classifier Using Speech Timing Information.}
The GPC models the relationship between input features and class labels within a probabilistic framework. Here, the input feature vector \( \mathbf{x} \in \mathbb{R}^d \) is constructed using the statistical characteristics (e.g., mean, variance) of durations from landmark bigrams identified as significant (\( p < 0.05 \)) in the earlier analysis, where \( d \) is the dimensionality of the feature vector. The class label \( y \in \{0, 1\} \) denotes either the depression or health category. 

The GPC predicts class probabilities \( P(y \mid \mathbf{x}) \) using a radial basis function (RBF) kernel:
\begin{equation}
k(\mathbf{x}, \mathbf{x}') = \sigma^2 \exp\left(-\frac{\|\mathbf{x} - \mathbf{x}'\|^2}{2l^2}\right)
\end{equation}
where \( \sigma^2 \) is the signal variance and \( l \) is the length scale, which controls the smoothness of the kernel function. This kernel facilitates capturing the non-linear relationships in the timing features, allowing robust probabilistic modeling of the speech data.

For a test feature vector \( \mathbf{x}_{\text{test}} \), the confidence score \( C \) is derived from the predicted probabilities as:

\small
\begin{equation}
C = \max(P(y = 0 \mid \mathbf{x}_{\text{test}}), P(y = 1 \mid \mathbf{x}_{\text{test}}))
\end{equation}
\normalsize

\paragraph{Expected Calibration Error (ECE)}
ECE is a widely used metric to evaluate the quality of confidence scores by measuring their alignment with observed accuracies~\cite{chaudhry2024finetuning, wu2024confidence}. For \( n \) test samples, ECE is computed as:
\begin{equation}
\text{ECE} = \sum_{b=1}^{B} \frac{|B_b|}{n} \left| \text{conf}(B_b) - \text{acc}(B_b) \right|
\end{equation}
where \( B_b \) denotes the set of samples in the \( b \)-th bin, and \( |B_b| \) is the number of samples in the bin. A smaller ECE value indicates better model calibration.

\section{Experiments}
\subsection{Experiments Setup}
\paragraph{Dataset}The DAIC-WOZ dataset~\cite{devault2014simsensei} is widely regarded as a benchmark resource for depression detection tasks. It comprises 189 recordings of clinical interviews conducted between interviewers and patients. Within the training set, 30 out of 107 interviews are labeled as depressed, while the development set contains 35 interviews, with 12 categorized as depressed. Following the practices of prior research, e.g.~\cite{gong2017topic,shen2022automatic,wu2022climate,wu2023self}, our experimental results are evaluated on the development set.

\paragraph{Model Configurations} We selected our evaluation models based on parameter efficiency and context processing capabilities. From the Llama2 family~\cite{touvron2023llama}, we included both base models (Llama2-7B, Llama2-13B) and their instruction-tuned variants (Llama2-7B Chat, Llama2-13B Chat) to assess the impact of instruction tuning. We also incorporated Llama3 models~\cite{dubey2024llama} (Llama3-8B-Instruct and Llama3-8B) for their enhanced context processing abilities. For Text-RAG baseline, we employed the SentenceTransformer~\cite{reimers-2019-sentence-bert} model \texttt{all-MiniLM-L6-v2} to compute embeddings and rank examples using cosine similarity, retrieving the top n examples (n = 1 or 2) from the training set.

\subsection{Main Results}
\begin{table}[t]
    \centering
    \setlength{\tabcolsep}{3pt} 
    \renewcommand{\arraystretch}{1.1} 
    \caption{F1 scores across different large language models and retrieval methods. The results for Speech Self-Supervised (SSL) models and Llama2 Fine-tune are taken from~\cite{wu2023self} and~\cite{zhang2024llms}, which applied extensive data augmentation specifically for depression detection. The "Method" column specifies the retrieval or input type, while the "Examples" column indicates the number of retrieved examples used.}
    \label{tab:llm_f1_scores}
    \scriptsize 
    \resizebox{\columnwidth}{!}{ 
    \begin{tabular}{@{}lcccccc@{}}
    \toprule
    \toprule
    \textbf{Model} & \textbf{Method} & \textbf{Examples} & \textbf{F1-avg} & \textbf{F1-max} & \textbf{F1-std} \\
    \midrule
    \midrule
    \textbf{Speech SSL Baselines} & Wav2Vec 2.0 & - & 0.627 & 0.667 & 0.043 \\
                                  & HuBERT & - & 0.667 & 0.762 & 0.052 \\
                                  & WavLM & - & 0.700 & 0.750 & 0.024 \\
    \midrule
    \textbf{Llama2 7B Chat} & Zero-shot & 0 & 0.195 & 0.207 & 0.023 \\
                            & Fine-tune & - & 0.488 & - & - \\
                            & Timing-RAG & 2 & 0.563 & 0.600 & 0.032 \\
    \midrule
    \textbf{Llama2 7B}      & Zero-shot & 0 & 0.173 & 0.182 & 0.011 \\
                            & Fine-tune & - & 0.578 & - & - \\
                            & Timing-RAG & 2 & 0.548 & 0.571 & 0.030 \\
    \midrule
    \textbf{Llama2 13B Chat} & Zero-shot & 0 & 0.186 & 0.191 & 0.005 \\
                            & Fine-tune & - & 0.545 & - & - \\
                            & Timing-RAG & 2 & 0.528 & 0.533 & 0.007 \\
    \midrule
    \textbf{Llama2 13B}     & Zero-shot & 0 & 0.249 & 0.300 & 0.037 \\
                            & Fine-tune & - & 0.636 & - & - \\
                            & Timing-RAG & 2 & 0.516 & 0.581 & 0.058 \\
    \midrule
    \textbf{Llama3 8B}      & Zero-shot & 0 & 0.528 & 0.537 & 0.008 \\
                            & Text-RAG & 1 & 0.458 & 0.487 & 0.020 \\
                            & Text-RAG & 2 & 0.292 & 0.316 & 0.038 \\
                            & Timing-RAG & 1 & 0.507 & 0.571 & 0.047 \\
                            & Timing-RAG & 2 & 0.601 & 0.640 & 0.026 \\
                            & Timing-RAG & 4 & 0.651 & 0.692 & 0.048 \\
    \midrule
    \textbf{Llama3 8B Instruct} & Zero-shot & 0 & 0.517 & 0.537 & 0.021 \\
                                & Text-RAG & 1 & 0.627 & 0.643 & 0.021 \\
                                & Text-RAG & 2 & 0.304 & 0.316 & 0.008 \\
                                & Timing-RAG & 1 & 0.497 & 0.500 & 0.007 \\
                                & Timing-RAG & 2 & 0.624 & 0.625 & 0.002 \\
                                & Timing-RAG & 4 & 0.692 & 0.733 & 0.049 \\
    \bottomrule
    \vspace{-8pt}
    \end{tabular}
    }
\end{table}

\begin{figure*}[t]
    \centering
    \begin{subfigure}{0.24\textwidth}
        \centering
        \includegraphics[width=\textwidth]{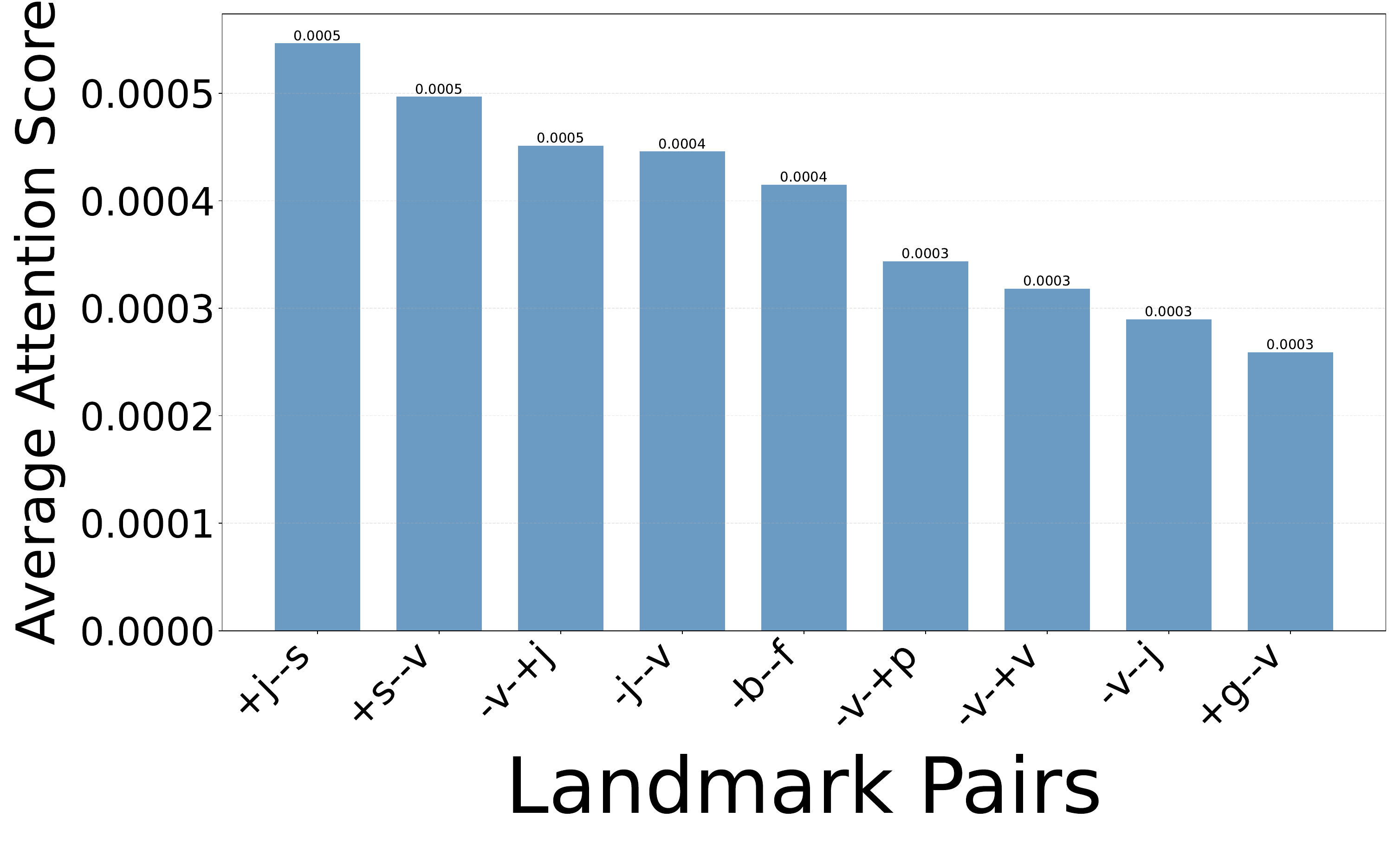}
        \caption{Early Layers (2-4)}
        \label{fig:landmark_early}
    \end{subfigure}%
    \hfill
    \begin{subfigure}{0.24\textwidth}
        \centering
        \includegraphics[width=\textwidth]{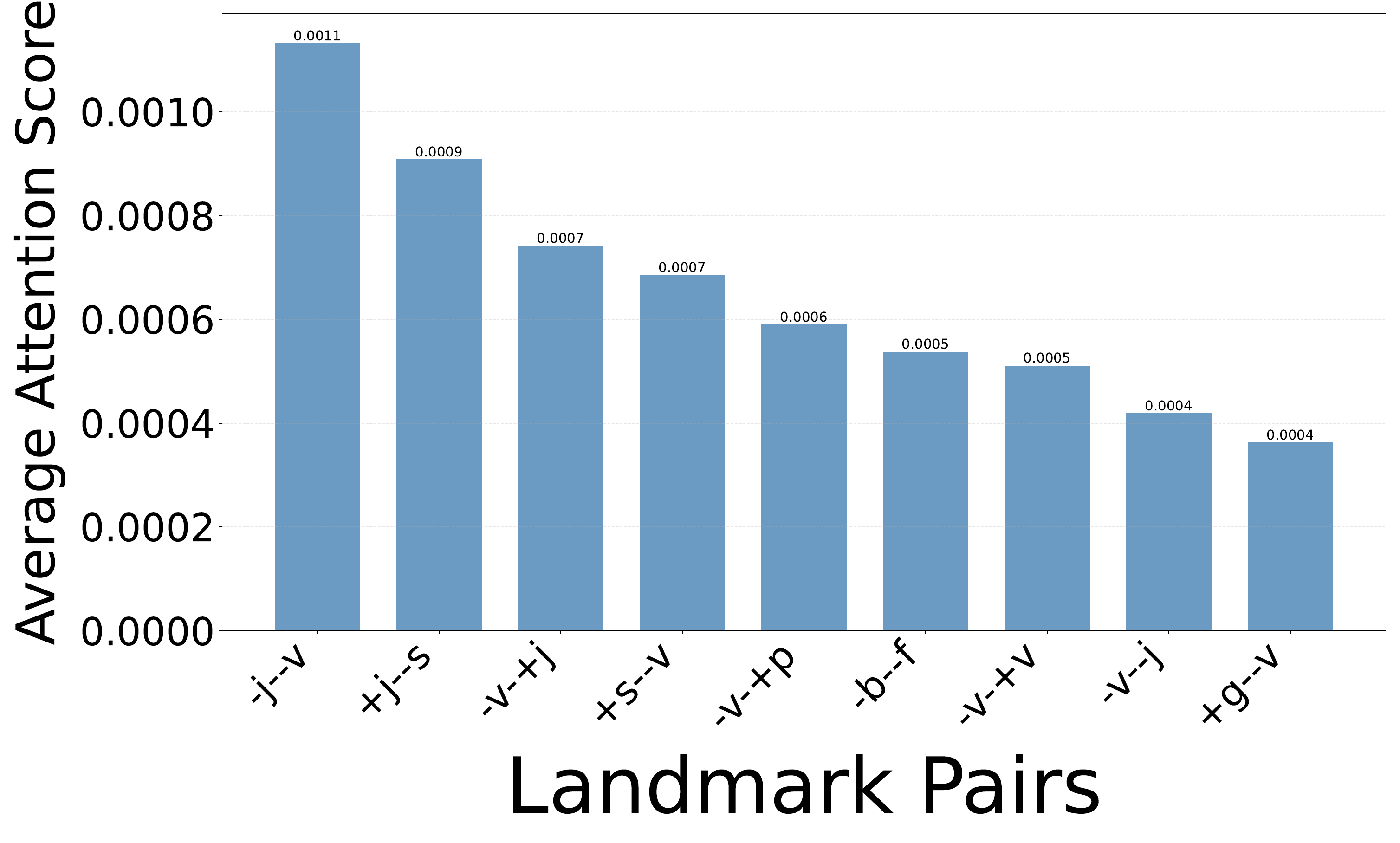}
        \caption{Middle Layers}
        \label{fig:landmark_middle}
    \end{subfigure}%
    \hfill
    \begin{subfigure}{0.24\textwidth}
        \centering
        \includegraphics[width=\textwidth]{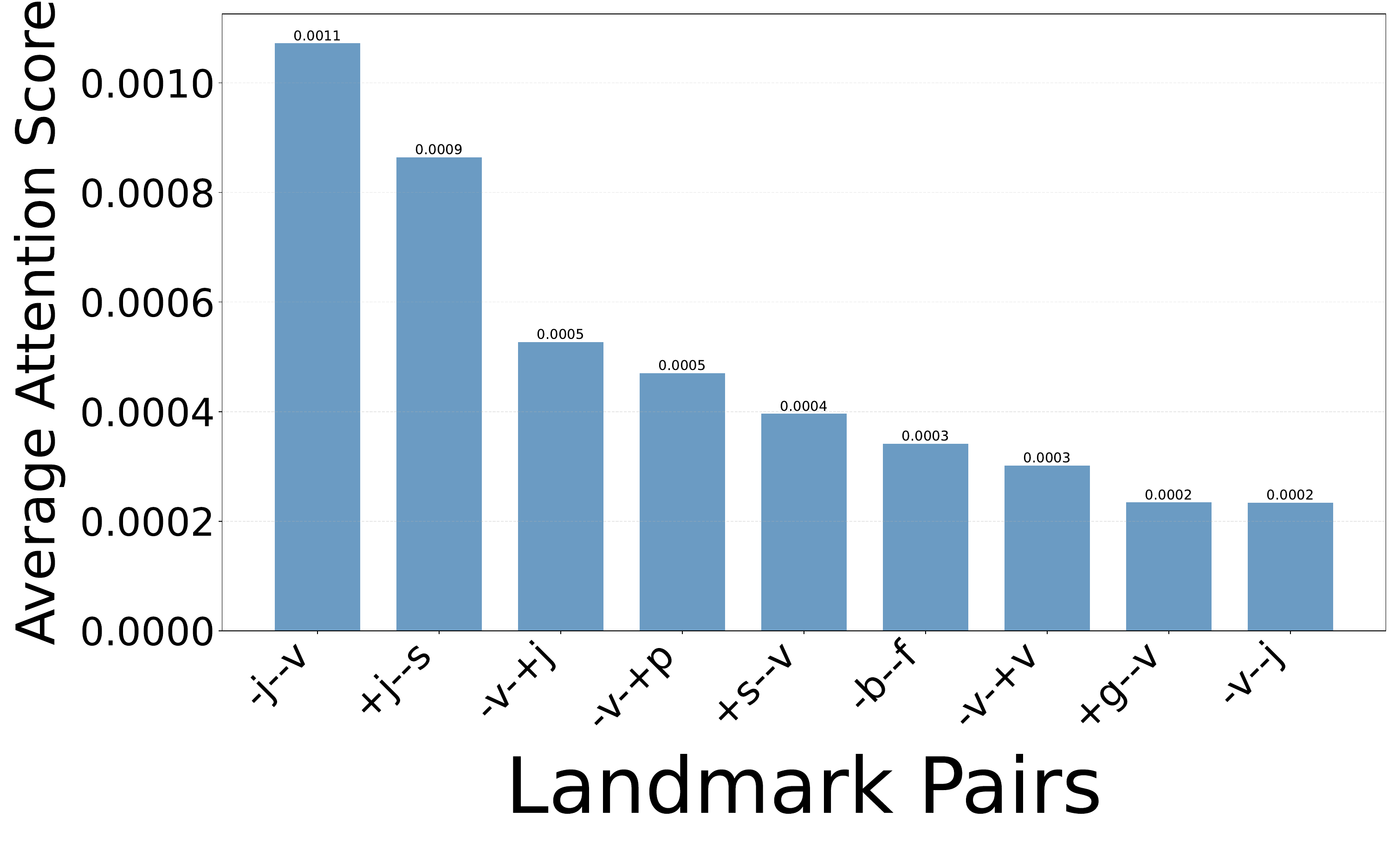}
        \caption{Final 4 Layers}
        \label{fig:landmark_final}
    \end{subfigure}%
    \hfill
    \begin{subfigure}{0.24\textwidth}
        \centering
        \includegraphics[width=\textwidth]{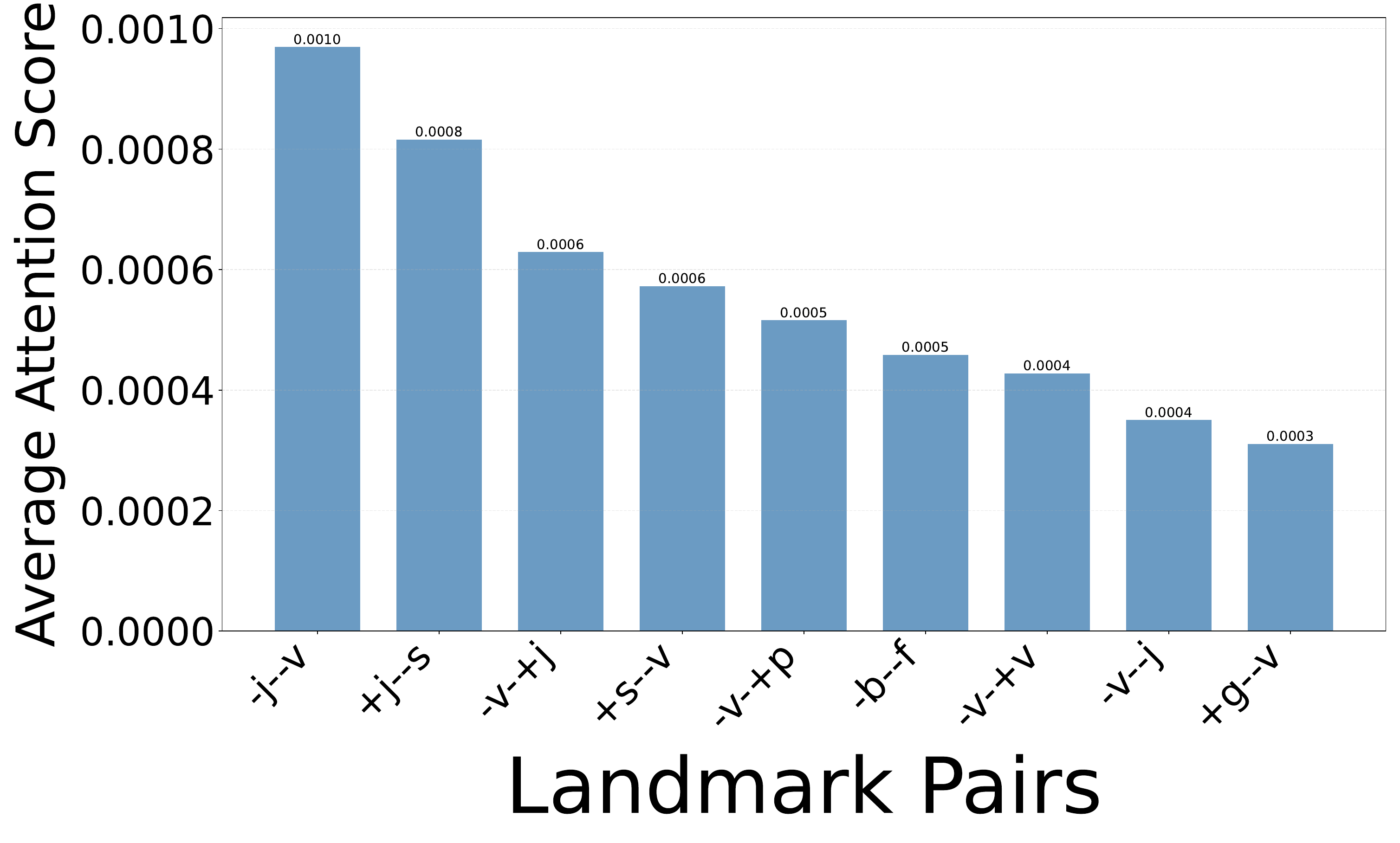}
        \caption{All Analyzed Layers}
        \label{fig:landmark_all}
    \end{subfigure}
    
    \caption{Attention scores across different landmark pairs for varying Transformer layers. Subfigures represent attention scores computed for early, middle, final, and all layers.}
    \vspace{-10pt}
    \label{fig:landmark_attention_scores}
\end{figure*}

Table~\ref{tab:llm_f1_scores} demonstrates the superior performance of our SpeechT-RAG approach over traditional Text-RAG methods in depression detection. As shown by the F1 scores, SpeechT-RAG exhibits consistent improvement with an increasing number of retrieved examples, highlighting the importance of temporal speech patterns in distinguishing between depressed and healthy individuals. In contrast, Text-RAG's performance deteriorates with additional retrieved examples, suggesting that text-based retrieval introduces noise that compromises classification accuracy.

The effectiveness of SpeechT-RAG is further validated across different model architectures. Despite Llama2's context window limitation allowing only two retrieved examples, it achieves performance comparable to fine-tuned models — establishing SpeechT-RAG as a resource-efficient alternative for scenarios with limited data or computational constraints. Leveraging Llama3's expanded context capacity, SpeechT-RAG demonstrates even stronger performance gains by effectively incorporating up to four examples, underscoring the scalability of our approach.

\begin{table}[t]
    \centering
    \setlength{\tabcolsep}{6pt} 
    \renewcommand{\arraystretch}{1.2} 
    \caption{Comparison of ECE results for Llama3 models and baselines. The results for MCDP~\cite{gal2016dropout} and PWLM~\cite{wu2024confidence} baselines are derived from~\cite{gal2016dropout} and~\cite{wu2023self}, respectively. The ECE values are calculated based on overall accuracy for the depression detection task. Mean ECE and standard deviation (Std) are reported.}
    \label{tab:ece_results}
    \scriptsize 
    \resizebox{\columnwidth}{!}{ 
    \begin{tabular}{@{}lcc@{}}
    \toprule
    \toprule
    \textbf{Model} & \textbf{Mean ECE (\textdownarrow)} & \textbf{ECE Std} \\
    \midrule
    \textbf{PWLM} & 0.183 & 0.009 \\
    \textbf{Llama3 8B MCDP} & 0.349 & 0.002 \\
    \textbf{Llama3 8B Instruct MCDP} & 0.340 & 0.003 \\
    \midrule
    \textbf{Llama3 8B} & 0.0674 & 0.0481 \\
    \textbf{Llama3 8B Instruct} & 0.0276 & 0.0274 \\
    \bottomrule
    \vspace{-8mm}
    \end{tabular}
    }
\end{table}
Table~\ref{tab:ece_results} presents the Expected Calibration Error (ECE) analysis, comparing our Timing-RAG approach with two established calibration methods. The first baseline, PWLM~\cite{wu2023self}, is a self-supervised speech model specifically trained for depression detection, with results cited from the original work. The second baseline, MCDP~\cite{wu2024confidence},employs Monte Carlo Dropout during prediction to estimate model uncertainty and was evaluated using both Llama3 8B and Llama3 8B Instruct models.

Our experimental results demonstrate the superior calibration performance of Timing-RAG. While MCDP achieves reasonable calibration with ECE values of 0.33 and 0.34, our Timing-RAG implementation with Llama3 8B and Llama3 8B Instruct models achieves substantially lower mean ECE values of 0.0674 and 0.0276, respectively. The consistently low standard deviations across trials further validate the reliability of our approach. These results underscore how incorporating speech timing information through Timing-RAG significantly enhances both the calibration quality and prediction reliability in LLM-based depression detection.

\section{Discussion}
\begin{figure*}[t]
    \centering
    \begin{subfigure}{0.24\textwidth}
        \centering
        \includegraphics[width=\textwidth]{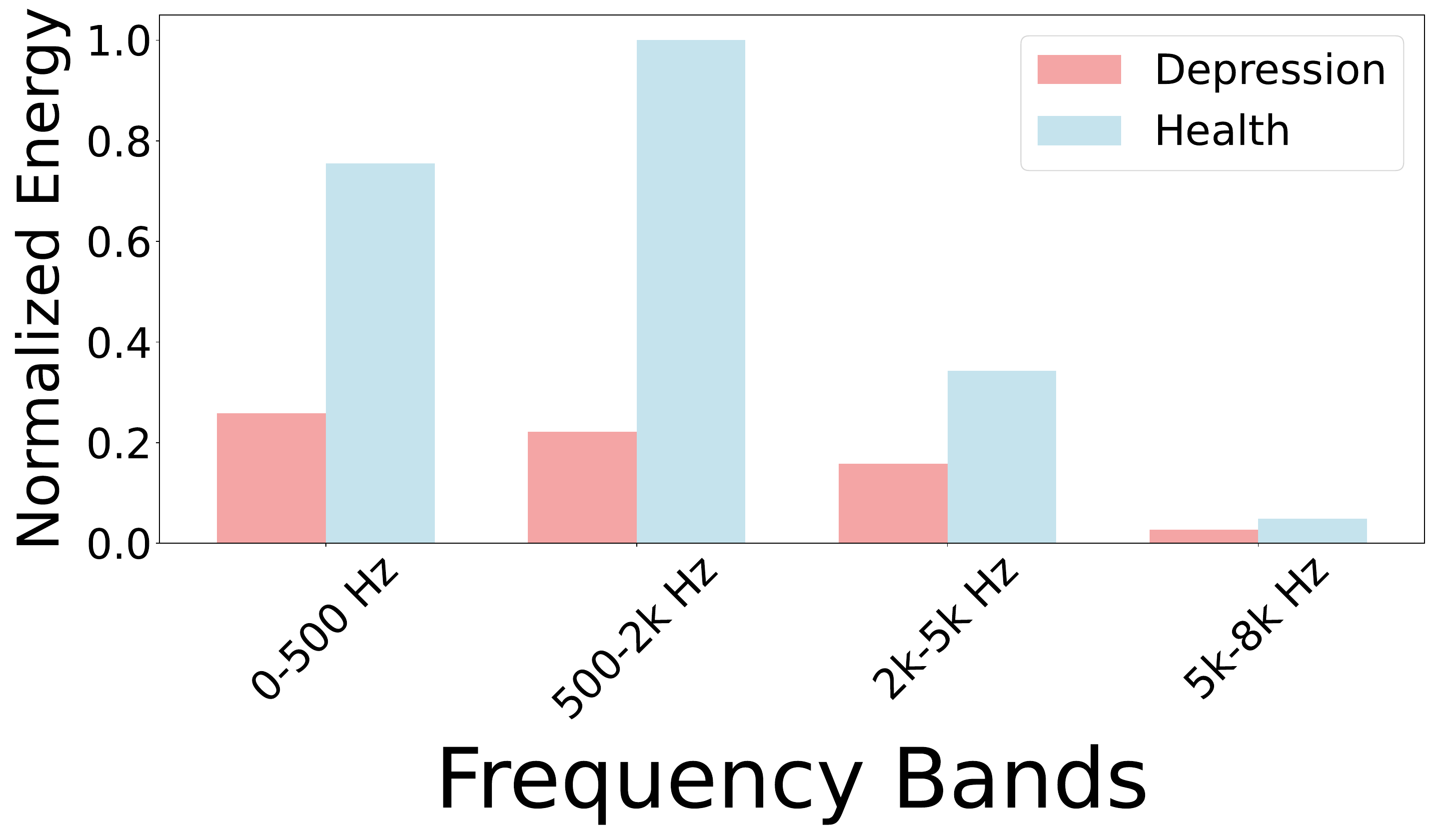}
        \label{fig:energy_bands}
    \end{subfigure}%
    \hfill
    \begin{subfigure}{0.24\textwidth}
        \centering
        \includegraphics[width=\textwidth]{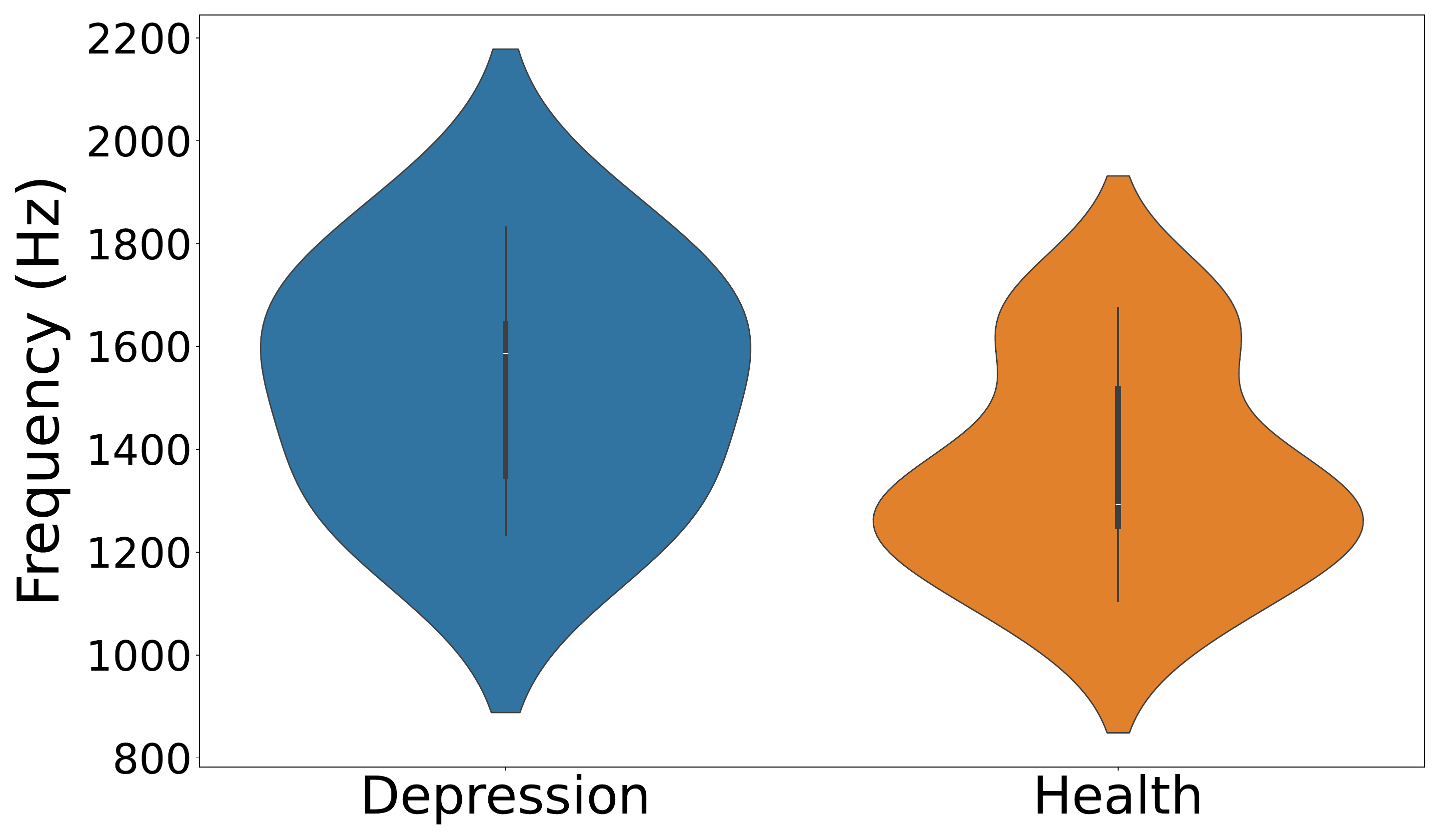}
        \label{fig:spectral_bandwidths}
    \end{subfigure}%
    \hfill
    \begin{subfigure}{0.24\textwidth}
        \centering
        \includegraphics[width=\textwidth]{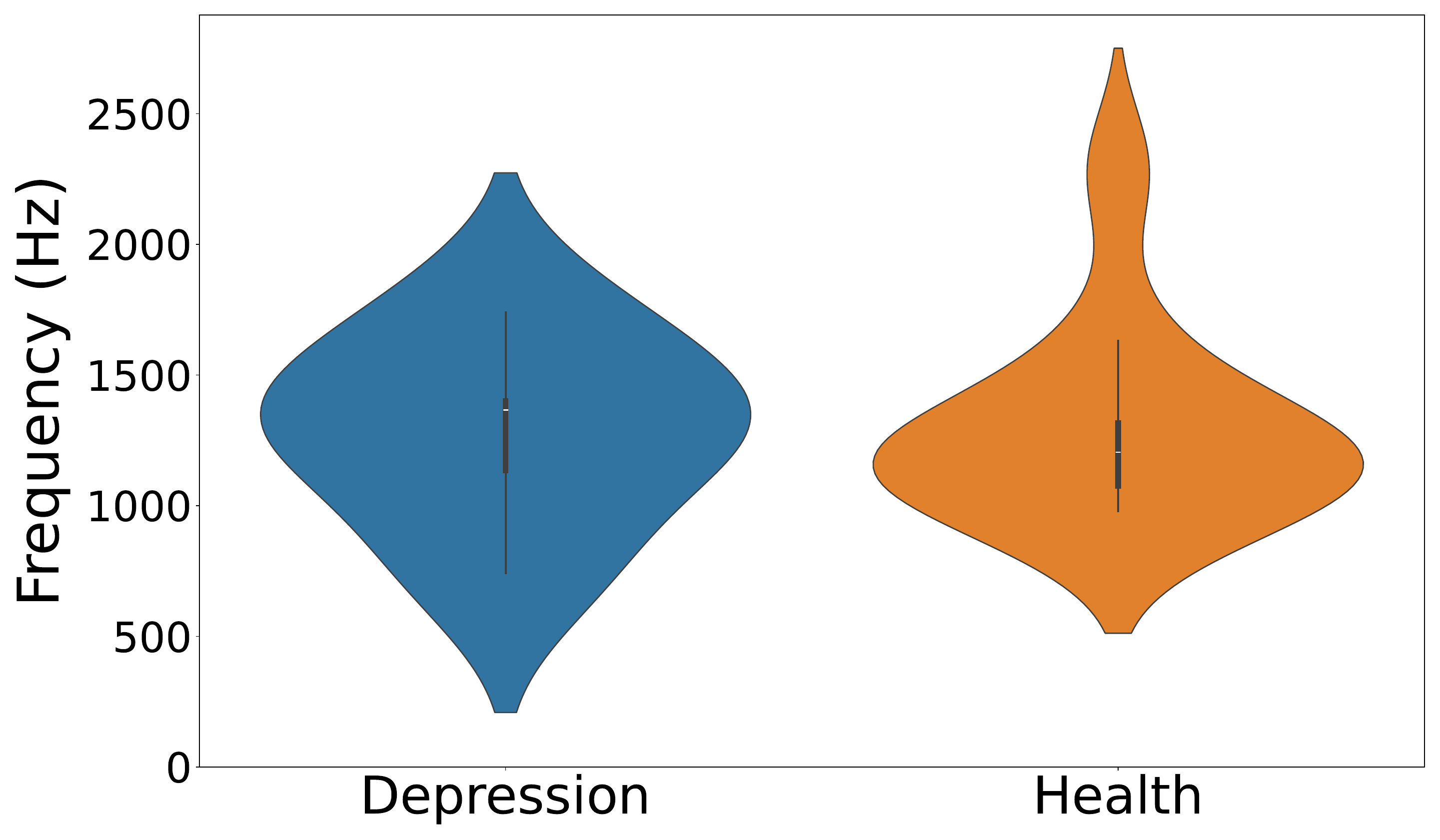}
        \label{fig:centroids}
    \end{subfigure}%
    \hfill
    \begin{subfigure}{0.24\textwidth}
        \centering
        \includegraphics[width=\textwidth]{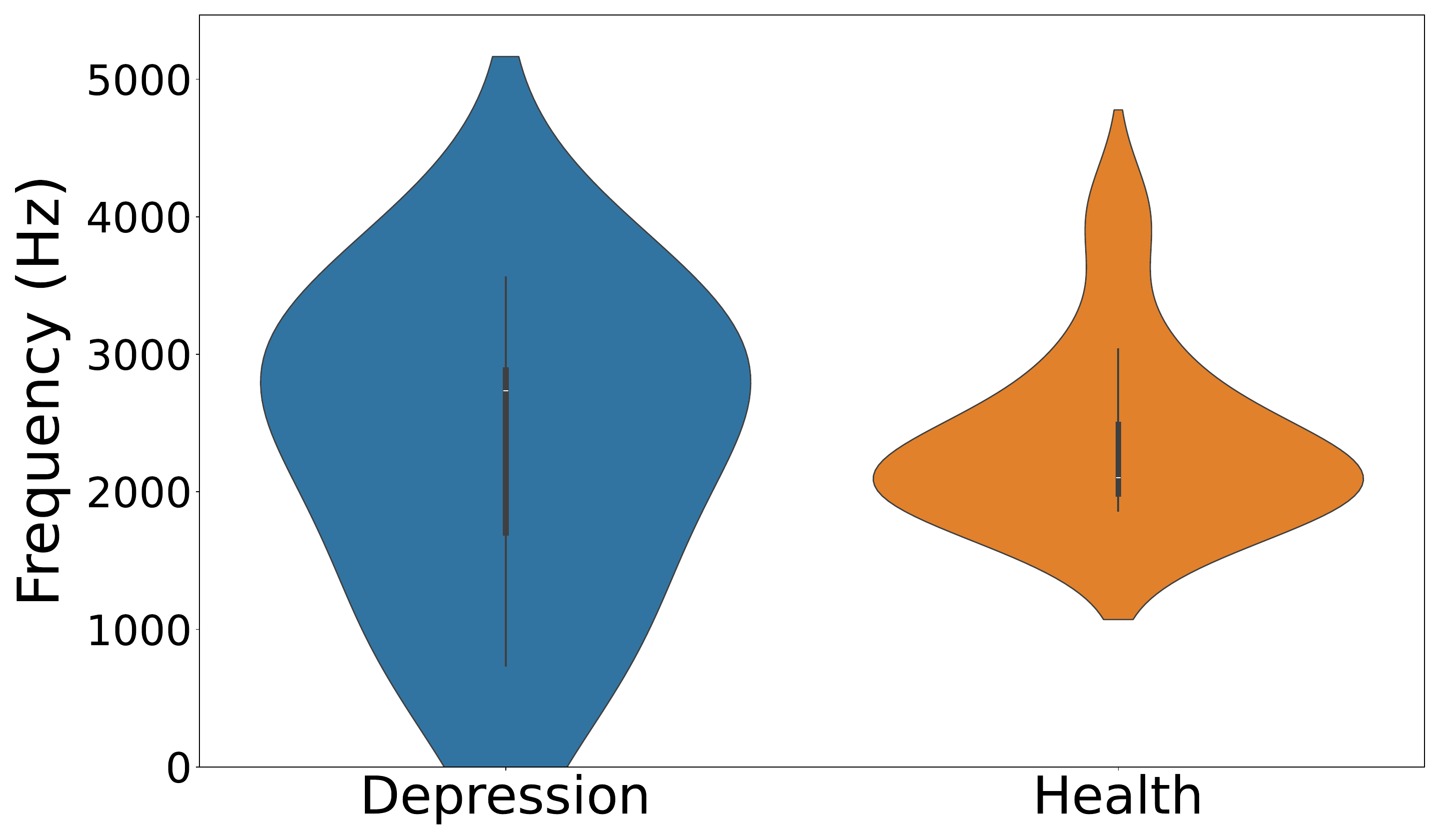}
        \label{fig:rolloffs}
    \end{subfigure}
    
    \caption{Acoustic analysis of speech segments for landmark pair (+s, -v): (a) Energy distribution across frequency bands; (b) Spectral bandwidth distribution; (c) Spectral centroid distribution; and (d) Spectral rolloff distribution, comparing depression and health groups.}
    \vspace{-10pt}
    \label{fig:Energy_Analysis}
\end{figure*}

\subsection{Landmark Pair Importance Analysis}
To gain a deeper understanding of how the Llama3 Instruct model leverages speech timing information for decision-making, we analyzed its attention mechanisms across different stages of processing. We selected the Llama3 Instruct model for this analysis based on its demonstrated stability and consistently high F1 scores across multiple experiments. The model's layers were categorized into three distinct groups: early layers (layers 2-4), middle layers (20\%-80\% of total layers), and final layers (the last 4 layers), enabling us to track the evolution of attention patterns throughout the model's processing pipeline.

For each layer group, we computed attention scores by identifying and aggregating attention weights assigned to tokens corresponding to landmark pairs in the input sequence. To quantify the importance of each landmark pair, we developed a scoring mechanism that considers both the magnitude and consistency of attention:
\begin{equation}
\text{Score} = \mu \times \left( 1 + 0.5 \cdot \sigma \right)
\end{equation}

where $\mu$ represents the mean attention score and $\sigma$ denotes the standard deviation across layers. This formulation extends beyond simple averaging by incorporating variance information, allowing us to distinguish between two scenarios: landmark pairs that maintain steady, high attention across layers (indicating consistent relevance to the model's decision process) versus those that receive sporadic high attention but may not be consistently meaningful. The scaling factor of 0.5 was empirically chosen to balance the influence of consistency versus absolute attention magnitude.

Building upon the earlier analysis of the model's attention mechanism, we observed that in the early layers, the Llama3 Instruct model prioritizes landmark pairs such as \( +s \to -v \) and \( -v \to +j \), which closely align with the most significant landmark pairs identified in our speech signal analysis, based on the lowest \( p \)-values. This similarity suggests that, similar to speech self-supervised learning (SSL) models~\cite{hsu2021hubert,pasad2021layer,chung2021similarity}, the Llama3 Instruct model may focus on low-level features in its early layers. This pattern suggests that both speech pre-trained models and text pre-trained models may first focus on low-level information before shifting to high-level information in their decision-making process for depression detection. In the middle and final layers, the Llama3 Instruct model focuses on similar landmark pairs, indicating a consistent attention pattern as the model processes speech features. These landmark pairs, such as \( +s \to -v \) and \( -v \to +j \),often associated with more abrupt changes in speech. The focus on these landmark pairs suggests that, in the deeper layers, the model emphasizes landmarks that represent more dynamic and notable speech features.

\subsection{Acoustic Validation of Detection-Critical Landmark Pairs}
Based on our analyses of both LLM attention patterns and statistical tests, we observed a consistent emphasis on the landmark pair (+s, -v), which appears to be particularly informative for depression detection. This finding motivates a detailed acoustic investigation of speech segments marked by this landmark pair.
As shown in Figure \ref{fig:Energy_Analysis}, we analyzed four key spectral features. The energy band distribution analysis reveals that healthy subjects maintain notably higher energy levels across all frequency bands, with the most pronounced difference in the speech fundamental frequency range (500Hz-2kHz). This difference suggests reduced vocal energy in depressed speech, particularly in frequencies crucial for speech articulation. For each band $b$, we calculate the normalized energy $E_b$ as:
\begin{equation}
E_b = \frac{1}{N} \sum_{f \in b} |X(f)|^2
\end{equation}
The spectral bandwidth distribution shows that depressed subjects exhibit a more compressed frequency spread, indicating less variability in their vocal frequency components. This reduced bandwidth suggests a more monotonic speech pattern, which aligns with clinical observations of reduced prosodic variation in depression~\cite{peper2012increase,quatieri2012vocal}.
The spectral centroid distribution further supports this finding, showing that depressed speech tends to have lower centroid values. This indicates that the center of mass of the frequency spectrum is shifted towards lower frequencies, suggesting a less "bright" or more dampened vocal quality characteristic of depressed speech.
The spectral roll-off analysis, representing the frequency below which 85\% of the spectral energy is contained, reveals that depressed speech consistently shows lower roll-off frequencies. This indicates a concentration of energy in lower frequency bands, further supporting the observation of reduced vocal expressiveness and energy in depressed speech patterns.

Collectively, these acoustic findings suggest that our landmark-based timing system captures specific energy patterns at critical acoustic transitions. The differences in energy distribution and frequency composition during the (+s, -v) segments demonstrate that these temporal transition points serve as meaningful acoustic indicators for depression detection.

\section{Conclusion}
In this paper, we have introduced SpeechT-RAG, a novel Retrieval-Augmented Generation framework that leverages acoustic temporal patterns for depression detection and uncertainty assessment. Through systematic analysis of LLM attention mechanisms and speech characteristics, we demonstrated how temporal information embedded in acoustic landmark pairs can simultaneously serve two critical functions: capturing depression-related patterns (such as reduced energy levels and diminished vocal expressiveness) and providing a natural basis for assessing prediction reliability. The dual utility of these temporal features enables our unified framework to not only achieve strong detection performance but also offer interpretable confidence estimation without additional computational overhead. These findings demonstrate how domain-specific temporal patterns can enhance both the accuracy and reliability of LLM frameworks, advancing the development of trustworthy systems.

\section*{Limitations}
A key limitation of our study is its reliance on the DAIC-WOZ dataset. While this dataset represents the current standard in multimodal depression recognition research and remains the only publicly accessible resource for speech-based depression analysis, this singular focus may impact the generalizability of our findings. The scarcity of available datasets in this domain stems from the significant privacy and ethical considerations surrounding mental health data collection. Nevertheless, our approach of conducting comprehensive analysis on this widely-used benchmark aligns with established research practices in the field of speech depression detection.
\section*{Ethics Statement}
Our research utilizes the DAIC-WOZ dataset, which has undergone rigorous de-identification procedures to ensure participant privacy protection. However, we acknowledge important ethical considerations regarding the deployment of our system. While our approach demonstrates improved accuracy in processing speech patterns for depression detection, the current performance level may not yet meet the stringent requirements for clinical applications. Like all machine learning models, our system may exhibit inherent biases that could lead to incorrect observations or classifications. Therefore, we emphasize that any practical implementation of this technology should be conducted under careful professional supervision, with our system serving as a supportive tool rather than a primary diagnostic mechanism.

\section*{Acknowledgement}
This work was supported by Australian Research Council Discovery Project DP230101184. 

\bibliography{reference}
\bibliographystyle{acl_natbib}

\appendix
\section{Example Prompts}
\label{sec:appendix}
\subsection{Zero Shot Prompt Example}
\begin{lstlisting}[basicstyle=\ttfamily\footnotesize, breaklines=true]
You are a mental health expert. 

Your task is to classify if a patient is depressed or healthy based on their dialogue.

You must respond with ONLY ONE WORD: either 'Depressed' or 'Health'.

Conversation:
{dialogue}

Diagnosis:

\end{lstlisting}
\subsection{Text Rag Prompt Example}

\begin{lstlisting}[basicstyle=\ttfamily\footnotesize, breaklines=true]
You are a mental health expert. 

Your task is to classify if a patient is depressed or healthy based on their dialogue.

You must respond with ONLY ONE WORD: either 'Depressed' or 'Health'.

"Case {idx}:
{example['dialogue']

Classification: {label}

Here are some example cases with their classifications:
{dialogue}

Now classify the following case with ONLY ONE WORD (Depressed or Health):

{dialogue}

Classification:

\end{lstlisting}

\subsection{Speech-Timing Rag Prompt Example}
\begin{lstlisting}[basicstyle=\ttfamily\footnotesize, breaklines=true]
"The task is to classify patients as 'Depression' or 'Health' based on their statistical feature patterns. "

"Each example below shows a sequence of statistical values followed by the correct classification Class."

format_example(item['bigram_durations_statics'], item['label'])

Class:

\end{lstlisting}
\end{document}